\documentclass[%
 reprint,
superscriptaddress,
 aps,prb,
]{revtex4-2}
\usepackage{graphicx}
\usepackage{dcolumn}
\usepackage{bm}
\usepackage{physics}
\usepackage{amsmath}
\usepackage[version=3]{mhchem}
\usepackage{subfig}
\usepackage{color}
\newcounter{num}
\newcommand{\Rnum}[1]{\setcounter{num}{#1}\Roman{num}}

\newcommand{\vk}{\vb*{k}}
\newcommand{\vq}{\vb*{q}}

\newcommand{\mean}[1]{\left\langle #1  \right\rangle}
\newcommand{\ep}{\varepsilon}
\newcommand{\gre}{\mathcal{G}}
\newcommand{\kbt}{k_B T}

\begin{document}

\preprint{APS/123-QED}

\title{Effect of paramagnon drag on thermoelectric transport properties: Linear response theory}% Force line breaks with \\

\author{\normalsize Junya Endo}
\affiliation{Department of Physics, University of Tokyo, Bunkyo, Tokyo 113-0033, Japan}

\author{\normalsize Hiroyasu Matsuura}%
\affiliation{Department of Physics, University of Tokyo, Bunkyo, Tokyo 113-0033, Japan}

\author{\normalsize Masao Ogata}
\affiliation{Department of Physics, University of Tokyo, Bunkyo, Tokyo 113-0033, Japan}
\affiliation{Trans-scale Quantum Science Institute, University of Tokyo, Bunkyo-ku, Tokyo 113-0033, Japan}

\date{\today}

\begin{abstract}
Paramagnetic materials exhibit a unique thermoelectric effect near the ferromagnetic transition point due to spin fluctuations. This phenomenon is referred to as paramagnon drag. In this study, we calculate the contribution of this paramagnon drag to the Seebeck coefficient microscopically based on the linear response theory. Consequently, we obtain a general formula for the contribution to the Seebeck coefficient due to the paramagnon drag, and then clarify the conditions in which the Seebeck coefficient is enhanced near the ferromagnetic transition point for a single-band and isotropic system. Moreover, we calculate the Seebeck coefficients for a mixture of free-electron-like and flat bands.
\end{abstract}

\maketitle

\section{Introduction}
Thermoelectric phenomena have attracted significant attention in recent years as an innovative technology that can extract energy from waste heat\cite{Sales1248,doi:10.1080/09506608.2016.1183075}. The Seebeck effect is a thermoelectric effect that produces a voltage difference $\Delta V$ from a temperature difference $\Delta T$. The Seebeck coefficient $S$ is defined as $S = -\Delta V / \Delta T$. Recently, attempts have been made to achieve high thermoelectric efficiency using magnetism\cite{Tsujiieaat5935,C6TA11120C,C8TC00788H,Zhengeaat9461,Hinterleitner2019}. From a theoretical perspective, the semiclassical Boltzmann transport theory is commonly used as a valid analytical method. However, in this method, it is difficult to include the effect of mutual interaction between electrons. \par
For microscopic analysis, the linear response to an external field applied to a thermal equilibrium state can be addressed using the Kubo formula\cite{doi:10.1143/JPSJ.12.570}. After its establishment, the formulation of the linear response theory based on the thermal Green's function\cite{Abrikosov} was a major breakthrough in the theoretical description of transport phenomena. In terms of thermoelectric phenomena, the Kubo formula was applied to calculate the thermal transport coefficients\cite{PhysRev.135.A1505}. \par
The effect of spin fluctuations on the thermoelectric effect in ferromagnets has been extensively studied both theoretically and experimentally\cite{PhysRev.126.2040,doi:10.1143/JPSJ.81.113602,PhysRevB.99.094425,Tripathi_2010,PhysRevB.94.144407}. In ferromagnets, spin waves "drag" electrons and cause a unique thermoelectric effect. The spin waves are quantized and described by quasi-particles, known as magnons; thus, this effect is called "magnon drag." In the magnon drag theory, the electron-magnon interaction is treated perturbatively, and the heat flow is described by calculating Green's function. A diagrammatic representation of the magnon drag has also been clarified\cite{doi:10.1143/JPSJ.81.113602,PhysRevB.99.094425}. Originally, the drag effect of the elementary excitation was proposed in the phonon case, which is called phonon drag\cite{Gurevich,doi:10.7566/JPSJ.88.074601}.\par
However, in the paramagnetic state, that is, above the magnetic transition temperature, some materials have been reported to exhibit a unique thermoelectric effect due to spin fluctuation. \ce{Fe2V_{0.9}Cr_{0.1}Al_{0.9}Si_{0.1}} is one of such materials, whose absolute value of the Seebeck coefficient decreases when a magnetic field is applied\cite{Tsujiieaat5935}. The origin has been proposed to be the suppression of the spin fluctuations in the magnetic field. Furthermore, in materials, such as \ce{RCo2} (R = Sc, Y, Lu) and \ce{AFe4Sb12} (A = Ca, Sr, Ba), a peculiar temperature dependence of the Seebeck coefficient has been observed, which is believed to be due to spin fluctuations\cite{Gratz_1995,GRATZ1997470,Gratz_2001,doi:10.1143/JPSJ.74.1382,TAKABATAKE200693}. The Seebeck effect due to spin fluctuation has been studied theoretically by Boltzmann theory, which introduces Kondo s-d coupling\cite{Okabe_2010}. In this study, the qualitative behavior of the Seebeck coefficient at low temperatures was investigated. \par
In the present paper, we discuss the Seebeck coefficient of the Hubbard model near the ferromagnetic transition point based on the linear response theory. We identify the dominant perturbation terms and heat flows, and calculate the quantitative dependence of the Seebeck coefficient on the energy dispersion of electrons. The Seebeck coefficient near the transition point is obtained quantitatively by considering the contribution of the Feynman diagrams corresponding to the paramagnon drag effect, referring to the diagrammatic representation of magnon drag. This results in a general expression for the correlation function due to the drag effect, assuming that the relaxation rate $\Gamma$ is small and constant. Because the theory is microscopic and rigorous, it applies to behavior at high temperatures. We apply this method to obtain transport coefficients in an isotropic system. In addition, we derive the conditions under which the Seebeck coefficient is enhanced by the paramagnon drag effect, and then show that the enhancement occurs in a mixture of free-electron-like and flat bands.

\begin{figure}[thpb]
    \subfloat[][]{\includegraphics[width=0.9\linewidth]{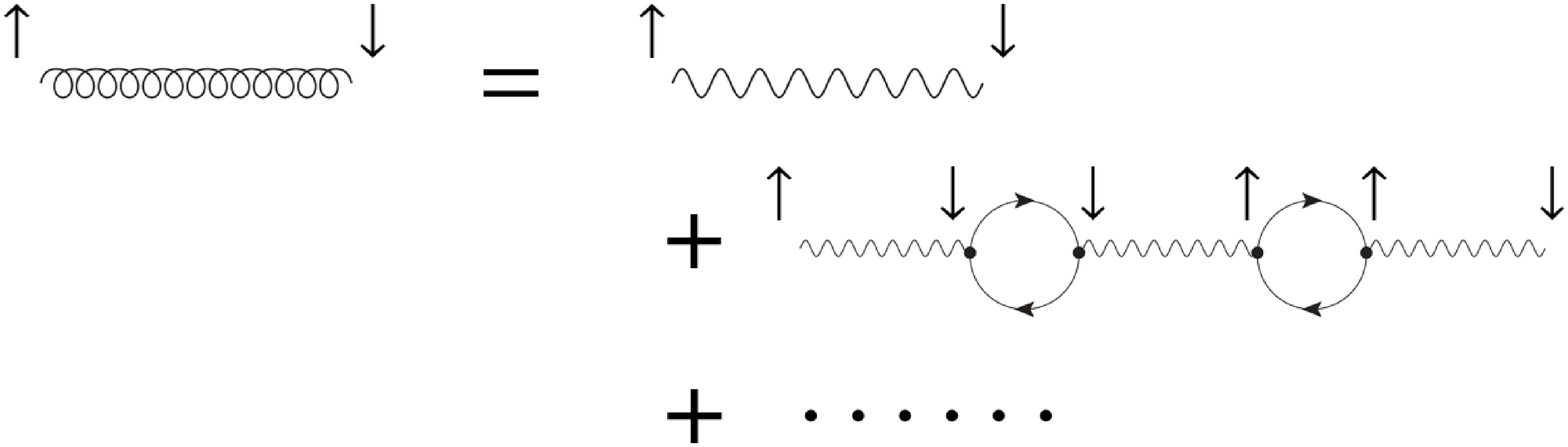}\label{subfig:int_rpa}}\\
    \subfloat[][]{\includegraphics[width=0.9\linewidth]{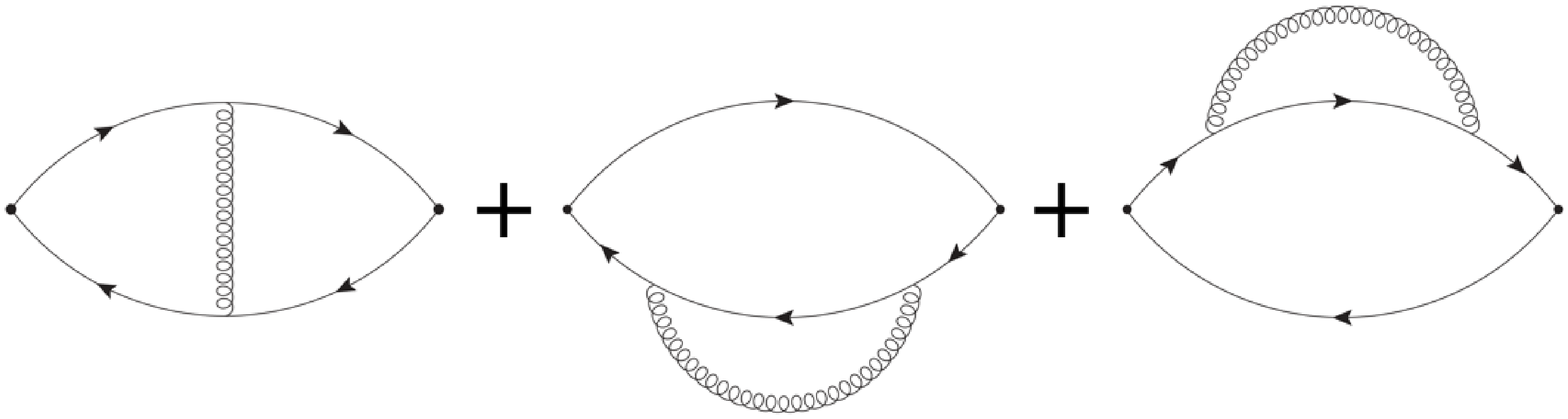}\label{subfig:MT}}\\
    \subfloat[][]{\includegraphics[width=0.9\linewidth]{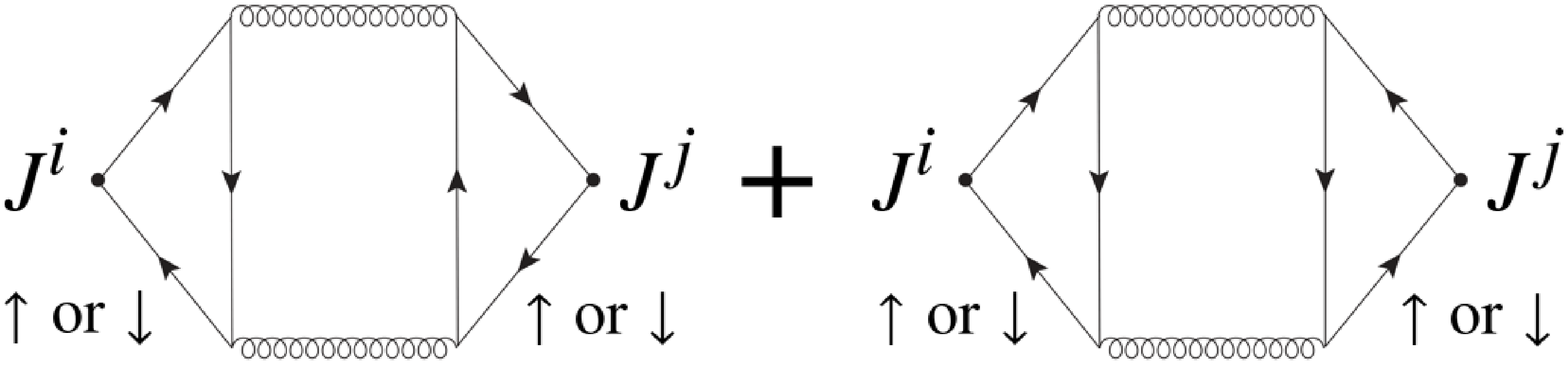}\label{subfig:drag}}
    \caption[]{\label{fig:drag} \subref{subfig:int_rpa} Feynman diagram of the ring approximation between up-spin and down-spin in the Hubbard model. For each loop, we take the sum of the Matsubara frequency and wavenumber. \subref{subfig:MT} Feynman diagrams of the first order of $U(\vq , \omega)$. These contributions are small (see Appendix \ref{section:appendix_first_U}). \subref{subfig:drag} Feynman diagrams for paramagnon drag effects.}
\end{figure}\par

\section{Paramagnon Drag contribution to transport coefficients}
To study the transport phenomena near the phase transition point from a paramagnetic state to a ferromagnetic state, we consider the ring approximation for the Hubbard interaction $U_{0}$. A diagram of this approximation is presented in Fig.~\ref{fig:drag}(a). The obtained effective interaction is related to Stoner's theory of ferromagnetism, and its divergence corresponds to the transition to ferromagnetism. The effective interactions can be written as:
\begin{eqnarray}
    U_{\uparrow \downarrow}(\vq , \omega) & = & \frac{U_{0}}{1-\pqty{U_{0} \chi(\vq, \omega)}^2}, \\
    U_{\uparrow \uparrow}(\vq , \omega) = U_{\downarrow \downarrow}(\vq , \omega) & = & \frac{U_{0}^{2}\chi(\vq, \omega)}{1-\pqty{U_{0} \chi(\vq, \omega)}^2},
\end{eqnarray}
where $\chi(\vq, \omega)$ denotes the dynamic susceptibility. $U_{0} \chi(\vb*{0}, 0) = 1$ is Stoner's criterion for the transition point. Because $U_{0} \chi \simeq 1$ is near the ferromagnetic transition point, we assume that $U_{\uparrow \downarrow}$ and $U_{\uparrow \uparrow}$ are equal for simplicity. \par
For further discussion, we considered two assumptions. First, the system is isotropic, and then $\ep_{\vk}$ depends only on $k = \abs{\vk}$. Second, $\ep_{k}$ is a monotonic function of $k$. Therefore, there is at most a single $k$ that satisfies $\ep_{k} = \ep_{F}$.
By performing a Taylor expansion of $\chi(\vq, \omega)$ around $q = 0$ and $\omega = 0$ under these assumptions, $U = U_{\uparrow \downarrow} = U_{\uparrow \uparrow}$ can be parameterized as 
\begin{equation}
    \label{eq:int_rpa}
    U^{R,A} (\vq , \omega) = \frac{U_{0}}{\eta + A \tilde{q}^2 \mp i C \tilde{\omega}/\tilde{q}},
\end{equation}
where $\tilde{q} = q/ k_F$, $\tilde{\omega} = \omega / \ep_F$\cite{doi:10.1143/JPSJ.18.1025}, and the superscript $R(A)$ represents the retarded (advanced) interaction that will be later used. In this parameterization, we assume that $\tilde{q},~\tilde{\omega}/\tilde{q} \ll 1$. As $\eta \to +0$, $U(\vb*{0} , 0)$ approaches infinity, which corresponds to a ferromagnetic transition. Simply, $\eta$ represents the distance from the critical point. Because we assume that the system is isotropic, the Taylor expansion $\abs{\vk - \vq} \simeq k - q\cos\theta + q^2 \sin^2 \theta / 2k$ for small $q$ can be used to generate Eq. \eqref{eq:int_rpa}. Using self-consistent renormalization (SCR) theory, we can determine the parameters $A$ and $C$ in a self-consistent manner\cite{doi:10.1143/JPSJ.34.639}. However, the specific values of these parameters are not important in the following discussion. \par
In this paper, we focus on the Seebeck coefficient $S = L^{12}/TL^{11}$, where $L^{ij}$ is the thermoelectric linear-response coefficient
\begin{equation}
    \vb*{J}^{1} = L^{11} \vb*{E} + L^{12} \pqty{- \frac{\nabla T}{T}}.
\end{equation}
Here, $\vb*{J}^{1}$, $\vb*{E}$, and $\nabla T$ are the electric current density, electric field, and temperature gradient, respectively.
The Kubo formula can be used to obtain $L^{ij}$. The correlation function is defined by
\begin{equation}
    \Phi^{ij}_{\mu \nu} (\vq , i\omega_{\lambda}) = \frac{1}{V} \int_{0}^{\beta} \dd \tau ~ \mean{\hat{J}^{i}_{H,\vq,\mu}(\tau)\hat{J}^{j}_{H,-\vq,\nu}(0)} e^{i\omega_{\lambda}\tau} ,
\end{equation}
where $V$ and $i\omega_{\lambda}$ are the volumes of the system and the Matsubara frequency, respectively. The subscript $H$ of $\hat{J}$ represents the Heisenberg representation, and $\mu,\nu$ represents the direction $x $ and $ y$. With the analytic prolongation of the Matsubsra frequency to the real frequency as $i\omega_{\lambda} \to \hbar \omega + i\delta$, $L^{ij}$ can be calculated as
\begin{equation}
    L^{ij} = \lim_{\omega \to 0} \frac{\Phi^{ij}_{xx}(\vb*{0} , \hbar\omega) - \Phi^{ij}_{xx}(\vb*{0} , 0)}{i(\omega + i \delta)}.
\end{equation}
\par
For the Hubbard model with energy dispersion $\ep_{\vk}$, the electric current density operator becomes
\begin{equation}
    \hat{J}^{1}_{\vq = \vb*{0},x} = \sum_{\vk , \sigma} ev_{\vk,x}c_{\vk\sigma}^{\dagger}c_{\vk\sigma},
\end{equation}
where $v_{\vk,x} = \partial\ep_{\vk}/\hbar\partial k_{x}$, and $e<0$. The heat current density operator is $\hat{J}^{2}_{\vq = \vb*{0},x} = \hat{J}^{2,kin}_{\vq = \vb*{0},x} + \hat{J}^{2,e-e}_{\vq = \vb*{0},x}$, where\cite{Mahan,PhysRevB.67.014408,doi:10.7566/JPSJ.88.074703}.
\begin{eqnarray}
    \hat{J}^{2,kin}_{\vq = \vb*{0},x} & = & \sum_{\vk , \sigma} (\ep_{\vk} - \mu)v_{\vk,x}c_{\vk\sigma}^{\dagger}c_{\vk\sigma}, \\
    \hat{J}^{2,e-e}_{\vq = \vb*{0},x} & = & \frac{1}{V} \sum_{\vk , \vk' , \vq' , \alpha\beta} U_{0} v_{\vk , x} \nonumber \\
    & ~ & \times c^{\dagger}_{\vk + \frac{\vq'}{2} ,\alpha}c^{\dagger}_{\vk' - \vq' , \beta} c_{\vk' , \beta} c_{\vk - \frac{\vq'}{2} , \alpha}.
\end{eqnarray}
Here, $\hat{J}^{2,kin}_{\vq = \vb*{0},x}$ is due to the kinetic energy of electrons, and $\hat{J}^{2,e-e}_{\vq = \vb*{0},x}$ is due to the electron-electron interactions. \par
As discussed in Ref.~\onlinecite{doi:10.7566/JPSJ.88.074703}, without the heat current of a long-range Coulomb interaction, a heat current of phonons, and a part of the heat currents occurring from the electron-phonon interaction, the following can be obtained
\begin{eqnarray}
    L^{11} & = & \int \dd \ep ~ (-f'(\ep)) \sigma(\ep, T),\\
    L^{12} & = & \frac{1}{e}\int \dd \ep ~ (-f'(\ep)) (\ep - \mu) \sigma(\ep, T),
\end{eqnarray}
where $\sigma (\ep , T)$ is the spectral conductivity, and here, it is referred to as the Sommerfeld-Bethe relation. Note that for the Hubbard model, we have $\hat{J}^{2,e-e}_{\vq = \vb*{0},x}$ , but the Sommerfeld-Bethe relation holds\cite{PhysRevB.67.014408}.
\par
To discuss the essence of the paramagnon drag effect, we consider the Feynman diagrams shown in Fig.~\ref{fig:drag}\subref{subfig:MT} (Maki-Thompson type), and Fig.~\ref{fig:drag}\subref{subfig:drag} (Aslamazov-Larkin type).
First, we show that the contribution of the first order of $U(\vq , \omega)$ in Fig.~\ref{fig:drag}\subref{subfig:MT} is smaller than that of the second order shown in Fig.~\ref{fig:drag}\subref{subfig:drag} in $\tilde{q} ,~ \tilde{\omega} \ll 1$, as shown in Appendix \ref{section:appendix_first_U}. Therefore, we consider that the main contribution of the paramagnon drag effect originates from the diagram in Fig.~\ref{fig:drag}\subref{subfig:drag}. In particular, in certain situations, it has been shown to represent the effects of spin fluctuations\cite{PhysRevB.94.115155,doi:10.1143/JPSJ.63.2042}. Therefore, the contributions of the diagram in Fig.~\ref{fig:drag}\subref{subfig:drag} are as follows.
\begin{widetext}
\begin{equation}
    \label{eq:phi_drag_m}
    \begin{split}
        \Phi^{ij}_{drag} (i\omega_{\lambda}) &= \frac{4(\kbt)^3}{V^3} \sum_{\vk_1, \vk_2, \vq} \sum_{l,m,n} \gre(\vk_1 , i\ep_{m-})\gre(\vk_1 , i\ep_{m})\gre(\vk_2 , i\ep_{n-})\gre(\vk_1 - \vq , i\ep_{m} - i\omega_{l})\gre(\vk_2 - \vq , i\ep_{n} - i\omega_{l}) \\
        &\times  U(\vq , i\omega_{l-}) U(\vq , i\omega_{l}) J^{i}_{x}(\vk_1) 
        \pqty{\gre(\vk_2 , i\ep_{n}) J^{j}_{x}(\vk_2) + \gre(\vk_2-\vq , i\ep_{n-}-i\omega_{l})J^{j}_{x}(\vk_2 - \vq)},
    \end{split}
\end{equation}
where factor 4 is derived from spin summations (see Fig.~\ref{fig:drag}\subref{subfig:drag}), $\ep_{m} = (2m+1)\pi\kbt$, and $\ep_{n} = (2n+1)\pi\kbt$ are fermion Matsubara frequencies, $\omega_{l} = 2l\pi\kbt$ is a boson Matsubara frequency, and $\ep_{m-} = \ep_{m} - \omega_{\lambda}$. 
We assume that the electron thermal Green's function satisfies $\gre\pqty{\vk, i\ep_{n}} = \pqty{i\ep_{n} - \ep_{\vk} + \mu + i\Gamma\mathrm{sign}(\ep_{n}) - \Sigma}^{-1}$, where $\Gamma$ is the relaxation rate of the electrons due to impurity scattering, and $\Sigma$ is the self-energy due to the spin fluctuation. In this study, we consider $\Gamma$ and neglect $\Sigma$ for simplicity. The effects of the real and imaginary parts of $\Sigma$ will be discussed qualitatively in Section \Rnum{4}.
Note that we have used $\hat{J}^{2,kin}_{\vq = \vb*{0},x}$ as the heat current operator; thus, $\hat{J}^{2}_{x} (\vk) = (\ep_{\vk} - \mu)v_{\vk x}$ and $\hat{J}^{1}_{x} (\vk) = ev_{\vk x}$. In Eq. \eqref{eq:phi_drag_m}, we reveal that the main contribution in the summation of $\omega_l$ is derived from the region of $0 < \omega_{l} < \omega_{\lambda}$, where $U(\vq , i\omega_{l-}) U(\vq , i\omega_{l})$ becomes $U^{A}U^{R}$ and the derivative of Bose distribution function $N'(\ep)$ appears. We consider the summation of $l,m,n$, and assume the analytic continuation $i\omega_{\lambda} \to \hbar \omega + i\delta$. In the summation of $m$ and $n$, the terms $G^{R}G^{R}G^{R}$ and $G^{A}G^{A}G^{A}$ can be neglected for $\Gamma / \ep_{F} \ll 1$. Here, $G^{R}$ and $G^{A}$ are the retarded and advanced Green's functions, respectively. In addition, we use the relation $G^{R}(\vk,\ep)G^{A}(\vk,\ep) \simeq \pi\delta(\ep - \ep_{\vk})/\Gamma$, and $G^{R}(\vk,\ep) - G^{A}(\vk,\ep) \simeq -2\pi i \delta(\ep - \ep_{\vk})$ for $\Gamma/\ep_{F} \ll 1$. Using these approximations, we obtain the following equation:
\begin{equation}
    \label{eq:phi_drag}
    \begin{split}
        \Phi^{ij}_{drag} (\hbar \omega + i\delta) &= -\frac{2\pi i \hbar \omega}{\Gamma^2 V^3} \sum_{\vk_1,\vk_2, \vq} N'(\Delta\ep_{1})\pqty{f(\ep_{\vk_1}) - f(\ep_{\vk_1 - \vq})}\pqty{f(\ep_{\vk_2}) - f(\ep_{\vk_2 - \vq})} \\
        &\times U^{R}(\vq,\Delta\ep_1)U^{A}(\vq,\Delta\ep_1)J^{i}_{x}(\vk_1) \pqty{J^{j}_{x}(\vk_2) - J^{j}_{x}(\vk_2-\vq)}\delta(\Delta\ep_1 - \Delta\ep_2) + O(\omega^2),
    \end{split}
\end{equation}
\end{widetext}
where $N(x)$ is the Bose distribution function $(e^{\beta x} - 1)^{-1}$, and $\Delta\ep_{i} = \ep_{\vk_i} - \ep_{\vk_i - \vq}$. It seems that the expression in (\ref{eq:phi_drag}) diverges when $\Delta \ep_{1} \to 0$ because $\abs{N'(x \to 0)} \to \infty$. However, as $\pqty{f(\ep_{\vk_1}) - f(\ep_{\vk_1 - \vq})}\pqty{f(\ep_{\vk_2}) - f(\ep_{\vk_2 - \vq})} \to 0$ when $\Delta\ep_1 = \Delta\ep_2 \to 0$, the divergence is merely a cosmetic singularity. \par
$U^{R}$ and $U^{A}$ become large, where $\abs{\vq} \ll k_{F}$ and $\abs{\Delta\ep_1} \ll \ep_{F},~ \kbt$. Therefore, we consider the Laurent expansion around $\vq = \vb*{0}$ and $\Delta\ep_1 = 0$. With this expansion, we obtain $N'(\Delta\ep_{1}) \simeq -\kbt / (\Delta\ep_{1})^2$. Using these expansions and integrating with respect to $\vk_{1}$ and $\vk_{2}$, the following can be obtained

\begin{widetext}
\begin{eqnarray}
    \label{eq:L11_drag}
    L^{11}_{drag} & \simeq & \frac{1}{V}\sum_{\vq}\frac{e^2 U_{0}^{2} \ep_{F} \tilde{q}_{x}^{2}}{16 \pi^2 \hbar \tilde{q} \Gamma^2} \frac{\kbt}{C(\eta + A \tilde{q}^{2})} \int \dd k_{2} ~ k_{2} (- f'(\ep_{k_{2}})) \int \dd k_{1} ~ k_{1} (- f'(\ep_{k_{1}})), \\
    \label{eq:L12_drag}
    L^{12}_{drag} & \simeq & \frac{1}{V}\sum_{\vq}\frac{e U_{0}^{2} \ep_{F} \tilde{q}_{x}^{2}}{16 \pi^2 \hbar \tilde{q} \Gamma^2} \frac{\kbt}{C(\eta + A \tilde{q}^{2})} \int \dd k_{2} ~ k_{2} (- f'(\ep_{k_{2}})) \int \dd k_{1} ~ k_{1} (\ep_{k_{1}} - \mu) (- f'(\ep_{k_{1}})).
\end{eqnarray}
\end{widetext}
\par
\begin{figure}[thpb]
    \includegraphics[width=0.5\linewidth]{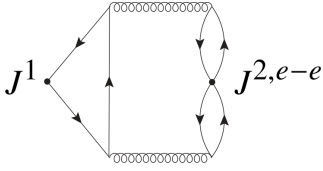}
    \caption{\label{fig:drag_ee} The lowest-order Feynman diagram between $J^{1} = \hat{J}_{\vq = 0 , x}^{1}$ and $J^{2,e-e} = \hat{J}_{\vq = 0 , x}^{2 , e-e}$, which corresponds to the paramagnon drag effect. This contribution can be neglected, as shown in Appendix \ref{section:appendix_e-e}.}
\end{figure}
The derivation of these results is presented in Appendix \ref{section:appendix_drag}. We can observe that Eqs. \eqref{eq:L11_drag} and \eqref{eq:L12_drag} satisfy the Sommerfeld-Bethe relation. \par
Note that for $L^{12}$, there are other contributions due to $\hat{J}^{2,e-e}_{\vq = 0, x}$. The lowest-order Feynman diagram using $\hat{J}_{\vq = 0 , x}^{2 , e-e}$, corresponding to the paramagnon drag effect, is shown in Fig.~\ref{fig:drag_ee}. However, we find that the contribution of this Feynman diagram is significantly small compared to the contributions of Fig.~\ref{fig:drag}(c) under the approximations of $\Gamma / \ep_{F} \ll 1$ and $\Delta \ep / \ep_{F} \ll 1$ (see Appendix \ref{section:appendix_e-e}). Thus, Eq. \eqref{eq:L12_drag} with only $\hat{J}^{2,kin}_{\vq = 0, x}$ satisfies the Sommerfeld-Bethe relation. Some approximations have been utilized in the derivation of Eqs. \eqref{eq:L11_drag} and \eqref{eq:L12_drag}. We can confirm that these approximations are valid for the case with $\eta \ll 1$ by computing the direct integration of Eq. \eqref{eq:phi_drag} by Monte Carlo integration with the replacement of $\delta(x) \simeq \Delta/\pi(x^2 + \Delta^2) \quad (\Delta \ll 1)$.

\section{Seebeck coefficient}
The Seebeck coefficient becomes
\begin{equation}
    \label{eq:seebeck_mix}
    S = \frac{1}{T}\frac{L^{12}_{free} + L^{12}_{drag}}{L^{11}_{free} + L^{11}_{drag}},
\end{equation}
where $L^{11}_{free}$ and $L^{12}_{free}$ are the contributions in the order of $\pqty{U_{0}}^0$,
\begin{eqnarray}
    L^{11}_{free} & = & \frac{\hbar}{\Gamma} \frac{1}{V} \sum_{\vk} e^2 v^2_{\vk,x} (-f'(\ep_{\vk})), \\
    L^{12}_{free} & = & \frac{\hbar}{\Gamma} \frac{1}{V} \sum_{\vk} e v^2_{\vk,x} (\ep_{\vk} - \mu) (-f'(\ep_{\vk})).
\end{eqnarray}
As $\eta$ approaches zero, that is, as the system approaches the ferromagnetic transition point, or as the temperature approaches the Curie temperature ($T_c$), we expect that $L^{ij}_{drag}$ dominates $L^{ij}_{free}$. In this case, $S$ can be expressed approximately as:
\begin{equation}
    \label{eq:S_drag_approx}
    S_{drag} = \frac{L^{12}_{drag}}{TL^{11}_{drag}} \simeq \frac{1}{eT} \frac{\int \dd k ~ k (\ep_{k} - \mu) (- f'(\ep_{k}))}{\int \dd k ~ k (- f'(\ep_{k}))}.
\end{equation}

However, as $\eta$ increases, $L^{ij}_{drag}$ becomes smaller than $L^{ij}_{free}$ and the Seebeck coefficient approaches
\begin{equation}
    \begin{split}
        S_{free} &= \frac{L^{12}_{free}}{TL^{11}_{free}} \\
        &= \frac{1}{eT} \frac{\int \dd k ~ k^2 \pqty{\pdv{\ep_{k}}{k}}^2 (\ep_{k} - \mu) (-f'(\ep_{k}))}{\int \dd k ~ k^2 \pqty{\pdv{\ep_{k}}{k}}^2 (-f'(\ep_{k}))}.
    \end{split}
\end{equation}

Therefore, the Seebeck coefficient gradually changes from $S_{free}$ to $S_{drag}$ as $T \to T_{c}$. 

\subsection{The condition for $\abs{S_{drag}} > \abs{S_{free}}$}

We can observe that the values of $S_{drag}$ and $S_{free}$ depend only on the functional form of $\ep_{k}$. It is difficult to determine the exact condition for $\abs{S_{drag}} > \abs{S_{free}}$ because they include integrals. However, when $\kbt / \mu \ll 1$, the condition can be achieved in a simple form using the Sommerfeld expansion. \par
We assume that $\ep_{k}$ is an increasing function of $k$, and $k_F$ is in the Brillouin zone. Thus, the Seebeck coefficients become 
\begin{eqnarray}
    \label{eq:S_free_somm}
    S_{free} & \simeq & -\frac{\pi^2 k_B^2 T}{3\abs{e}} \frac{2\ep'(k_F) + k_F \ep''(k_F)}{k_F (\ep'(k_F))^2}, \\
    \label{eq:S_drag_somm}
    S_{drag} & \simeq & -\frac{\pi^2 k_B^2 T}{3\abs{e}} \frac{\ep'(k_F) - k_F \ep''(k_F)}{k_F (\ep'(k_F))^2}.
\end{eqnarray}
Here, we used the Sommerfeld expansion after changing the wavenumber integral to the energy integral as $k=k(\ep)$.
For example, in the case of $\ep_{k} \propto k^{2}$ (free electron), $S_{drag}/S_{free} \simeq 0$. Because we assume that $\ep'(k_{F}) > 0$, $\abs{S_{drag}} > \abs{S_{free}}$ holds only when $k_{F} \ep'(k_{F}) < -\ep'(k_{F})/2$.
Thus, the dispersion relation should be upwardly convex.

\begin{figure}
    \includegraphics[width=0.9\linewidth]{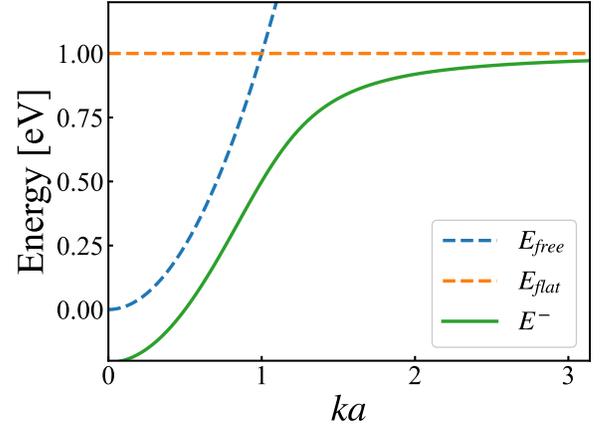}\\
    \caption{\label{fig:band} Energy dispersion $E^{-}$ in comparison with that of the free electron band.} 
\end{figure}

\subsection{A model with a flat band and free electron band}
As an example, we study a two-band model that realizes $\ep''(k_{F}) < 0$. The kinetic energy part of the Hamiltonian is expressed by:
\begin{equation}
    \mathcal{H} = \mqty(\hbar^2 k^2 / 2m & V \\ V^{*} & E_0),
\end{equation}
where $E_0 (>0)$ is the energy of the flat band, and $V$ represents the mixing term between the dispersive electron band and the flat band. The eigenenergies of this Hamiltonian are 
\begin{equation}
    E^{\pm} = \frac{1}{2}\pqty{K^2 + E_0 \pm \sqrt{(E_0 - K^2)^2 + 4\abs{V}^2}},    
\end{equation}
where $K^2 = \hbar^2 k^2 / 2m$. For simplicity, assume that the chemical potential satisfies $0 < \mu < E_0$ and consider only the band corresponding to $E^{-}$. This model can be used as an effective model for the present work because the shape of the band, similar to the flat band, increases the density of states and facilitates the ferromagnetic transition.\par
\begin{figure}
    \includegraphics[width=0.9\linewidth]{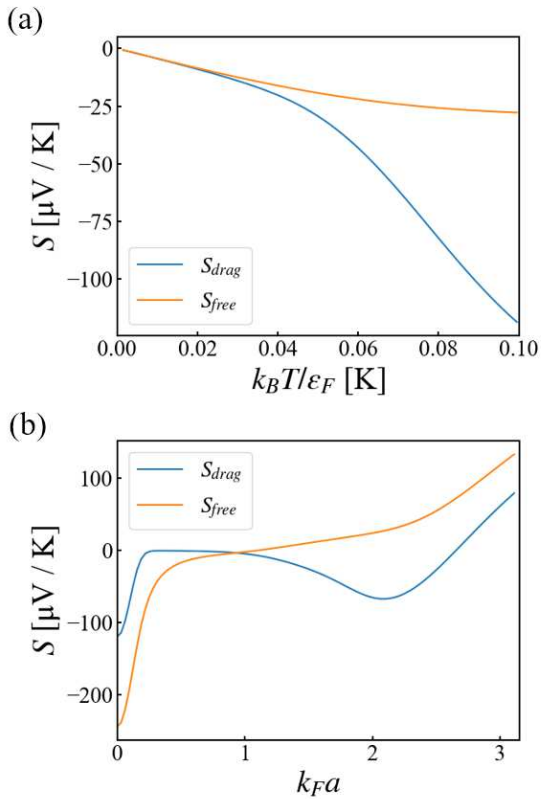}\\
    \caption[]{\label{fig:flat} (a) Temperature dependence of the Seebeck coefficient of $E^{-}$ band due to drag effect $(S_{drag})$, and one at $U=0$ $(S_{free})$ with $k_{F}a = 0.93$. At this $k_{F}$, $S_{drag} \simeq S_{free}$ at low temperatures. (b) $k_{F}$ dependence at $T = 100 ~ \mathrm{K}$.}
\end{figure}

On the basis of this model, we calculate the Seebeck coefficient $S_{drag}$ due to the paramagnon drag effect induced by the Hubbard interaction. We numerically perform the integration of Eqs. \eqref{eq:L11_drag} and \eqref{eq:L12_drag}, and evaluate $S_{drag} = L_{drag}^{12}/TL_{drag}^{11}$, in the parameter set of $E_{0} = 1.0~\mathrm{eV}$, $V = 0.5~\mathrm{eV}$, and $\hbar^2 / 2ma^2 = 1.0~\mathrm{eV}$, where $a$ is the lattice constant. With these parameters, the energy dispersion $E^{-}$ is shown in Fig.~\ref{fig:band}, and the Seebeck coefficients obtained are shown in Fig.~\ref{fig:flat}. Figure \ref{fig:flat}(a) shows the temperature dependence of $S_{drag}$ and $S_{free}$ for $k_{F}a = 0.93$. At low temperatures, the Seebeck coefficients are proportional to the temperature, and $S_{free} \simeq S_{drag}$ is satisfied because we chose $k_{F}$ that satisfies $\ep'(k_F) + 2k_F \ep'(k_F) \simeq 0$. At high temperatures, it can be observed that $S_{drag}$ is larger than $S_{free}$ at high temperatures. This is because the effect of the energy dispersion away from the Fermi surface starts to contribute. Note that because the actual Seebeck coefficient $S$ can be expressed as in Eq. \eqref{eq:seebeck_mix}, and therefore $S_{drag} < S < S_{free}$ or $S_{free} < S < S_{drag}$. It is considered that $S \simeq S_{drag}$ near the Curie temperature. Figure \ref{fig:flat}(b) shows the dependence of $S_{drag}$ and $S_{free}$ on the Fermi wavenumber. $-S_{drag}$ becomes larger than $-S_{free}$ in the region of $k_{F}a > 0.93$, where the energy dispersion $E^{-}$ is convex upward. In this region, $-S$ is expected to increase as the temperature approaches $T_{c}$ because $-S_{drag} > -S > -S_{free}$ is satisfied and $S$ approaches $S_{drag}$ as $T \to T_{c}$. The sign change of the Seebeck coefficient at large $k_{F}$ is caused by considering only a single band and restricting the integration only in the Brillouin zone.

\section{Discussion and Conclusion}
Finally, we discuss the additional contribution to the Seebeck effect from the self-energy due to the spin fluctuation $\Sigma$. Firstly, $\mathrm{Im}~\Sigma(0)$ does not affect $S_{drag}$ because of the cancellation similar to the cancellation of the constant $\Gamma$ in $L^{12}/L^{11}$. Secondly, when we consider the $\omega$ dependence of $\mathrm{Im}~\Sigma$ and the Sommerfeld-Bethe relation, an additional contribution proportional to $\dd \log \pqty{\mathrm{Im}~\Sigma}/\dd \omega$ will appear in the Seebeck coefficient. However, it was shown that $\mathrm{Im}~\Sigma$ is an even function of $\omega$ near the Fermi energy\cite{PhysRevB.37.3299,Schossmann1987}, and thus we can approximate the self-energy as $-\mathrm{Im}~\Sigma \simeq \Gamma_0 + a\omega^{2}$. In this case, the amplitude of $\dd \log \pqty{\mathrm{Im}~\Sigma}/\dd \omega$ is expected to be smaller than that in Eq. \eqref{eq:S_drag_approx}. As for the real part of the self-energy, the change as a function of $k$ around $k = k_{F}$ is found to be sufficiently slow\cite{PhysRevB.37.3299}. Therefore, the shape of the band dispersion cannot change significantly, and the real part of the self-energy simply renormalizes the effective mass and chemical potential.\par
These arguments are qualitative, and we need a quantitative discussion to obtain more precise results for the self-energy due to the spin fluctuation. This can be achieved by considering Feynman's diagrams we neglected in this study and by examining the details of the self-energy for more specific models. However, it is beyond our approximation, and this remains a future problem.\par
In conclusion, we investigated the behavior of the Seebeck coefficient near the ferromagnetic transition temperature in the Hubbard model, considering the paramagnon drag effects. In particular, for an isotropic system, $S_{drag}$ was calculated using Eqs. \eqref{eq:L11_drag} and \eqref{eq:L12_drag}. We found the condition of $\ep_{k}$, where the contribution of the paramagnon drag is large. In addition, as a model for a large Seebeck coefficient near $T_{c}$, we performed specific calculations for a mixture of free-electron-like and flat bands. 
We showed that the paramagnon drag effect can enhance the Seebeck coefficient near the ferromagnetic transition temperature. The present method provides a basis for the prediction of materials with large paramagnon drag contributions.

\section*{Acknowledgement}
We are grateful to T. Mori and N. Tsujii for fruitful discussions. This work was supported by Grants-in-Aid for Scientific Research from the Japan Society for the Promotion of Science (Nos. JP20K03802, JP18H01162, and JP18K03482), and the JST-Mirai Program Grant Number JPMJMI19A1, Japan. J. Endo was supported by the Japan Society for the Promotion of Science through the Program for Leading Graduate Schools (MERIT).

\appendix

\begin{widetext}

\section{The first order of $U$}
\label{section:appendix_first_U}

The contribution of the first order of $U(\vq , \omega)$ shown in Fig.~\ref{fig:drag}\subref{subfig:MT} is
    \begin{equation}
        \begin{split}
            \Phi^{ij,(1)}_{xx} (i\omega_{\lambda}) &= \frac{2(\kbt)^2}{V^2}\sum_{l,n} \sum_{\vk , \vq}  \gre(\vk , i\ep_{n-}) \gre(\vk , i\ep_{n}) J^{i}_{x}(\vk) \left\{\gre(\vk-\vq , i\ep_{n-}-i\omega_{l}) \gre(\vk - \vq , i\ep_{n} - i\omega_{l}) J^{j}_{x}(\vk - \vq) \right.\\
            &~~+ \left.  \pqty{\gre(\vk , i\ep_{n}) \gre(\vk - \vq , i\ep_{n} - i\omega_{l}) + \gre(\vk , i\ep_{n-}) \gre(\vk - \vq , i\ep_{n-} - i\omega_{l}) }J^{j}_{x}(\vk) \right\} U(\vq , i\omega_{l}).
        \end{split}
    \end{equation}
    Considering the summation of the Matsubara frequencies and assuming $\Delta \ep = \ep_{\vk} - \ep_{\vk - \vq} \ll \ep_{F}$, we obtain
    \begin{equation}
        \label{eq:first_U}
        \begin{split}
            \Phi^{ij,(1)}_{xx} (\hbar \omega + i\delta) &= \frac{\hbar\omega}{2\Gamma^2} N'(\Delta\ep) \pqty{f(\ep_{\vk}) - f(\ep_{\vk - \vq})} \pqty{U^{R}(\vq , \Delta\ep) - U^{A}(\vq , \Delta\ep)} J^{i}_{x}(\vk) \pqty{J^{j}_{x}(\vk) - J^{j}_{x}(\vk - \vq)}.
        \end{split}
    \end{equation}
    We used the same approximation as in the text; that is, $G^{R}G^{R}G^{R}G^{R}$ and $G^{A}G^{A}G^{A}G^{A}$ were neglected, and $-\omega_{\lambda} < \omega_{l} < \omega_{\lambda}$ was assumed.
    When we calculate Eq. \eqref{eq:first_U} in the region of $\abs{\vq} \ll k_{F}$, we find that it is proportional to $\tilde{q}^3$, which is in the higher order than Eq. \eqref{eq:phi_drag}. Therefore, this contribution is smaller than that shown in Fig.~\ref{fig:drag}\subref{subfig:drag}.

\section{The derivation of Eqs. \eqref{eq:L11_drag} and \eqref{eq:L12_drag}}
\label{section:appendix_drag}
First, assuming that $\Delta \ep_{1} = \Delta \ep_{2} \ll \ep_{F}$, Eq. \eqref{eq:phi_drag} divided by $i\omega$ becomes
\begin{equation}
    \frac{2\pi \hbar \kbt}{\Gamma^2 V^3} \sum_{\vk_{1},\vk_{2},\vq}f'(\ep_{\vk_1})f'(\ep_{\vk_2})\frac{U_0^2}{\pqty{\eta + A\tilde{q}^2}^2 + \pqty{C \tilde{\ep}_{\vk_{1}}/\tilde{q}}^2} 
    \times J_{x}^{i}(\vk_{1}) \pqty{J_{x}^{j}(\vk_{2}) - J_{x}^{j}(\vk_{2} - \vq)} \delta\pqty{\Delta \ep_{1} - \Delta \ep_{2}}.
\end{equation}
$L^{12}_{drag}$ can be calculated in the same manner. In an isotropic system, 
\begin{eqnarray}
    \Delta \ep_{i} & \simeq & \pqty{q\cos\theta_{i} - \frac{q^{2}}{2k_{i}}} \ep'(k_{i}), \\
    J_{x}^{1}(\vk_{1}) & \simeq & \frac{\cos\theta_{1} \cos\alpha + \sin\theta_{1} \sin \varphi_{1} \sin\alpha}{\hbar} e \ep'(k_{1}), \\
    J_{x}^{1}(\vk_{2}) - J_{x}^{1}(\vk_{2} - \vq) & \simeq & \frac{q \cos\alpha}{\hbar k_{2}} e \ep'(k_{2}).
\end{eqnarray}
where $\theta_{i}$ is the angle between $\vk_{i}$ and $\vq$, and $\alpha$ is the angle between the $\vq$ and $x$ directions. After the integral of $\theta_{1}, \theta_{2}, \varphi_{1}$, and $\varphi_{2}$, we obtain
\begin{equation}
    \begin{split}
        L^{11}_{drag} &\simeq \frac{\kbt}{\pqty{2\pi}^3 \hbar \Gamma^2 V} \sum_{\vq} \cos^2 \alpha \int \dd k_{2} ~ k_{2} (-f'(\ep(k_{2}))) \\
        &\times \int \dd k_{1}~k_{1}^2 \ep'(k_1) \frac{1}{\eta + A \tilde{q}^2} \frac{\tilde{q}}{C k_{1} \tilde{\ep}'(k_{1})} \mathrm{Arctan} \pqty{\frac{\pqty{1 - \tilde{q}/2\tilde{k}_{1}}C\tilde{q}\tilde{\ep}'(k_{1})}{\eta + A\tilde{q}^2}} (-f'(\ep(k_{1}))).
    \end{split}
\end{equation}
By approximating $\mathrm{Arctan}(x)$ as a constant, $\pi / 2$, we obtain Eq. \eqref{eq:L11_drag}.

\section{The contribution of $J^{2,e-e}_{\vq=0,x}$}
\label{section:appendix_e-e}
To certify that the contribution of $J^{2,kin}_{\vq=0 , x}$ is dominant, our calculation are presented in Fig.~\ref{fig:drag_ee}. Writing down the diagram in the form of thermal Green's functions, we obtain
    \begin{equation}
        \label{eq:phi_drag_m_e-e}
        \begin{split}
            \Phi^{12}_{e-e} (i\omega_{\lambda}) &= \frac{8(\kbt)^4}{V^4} \sum_{\vk_1, \vk_2, \vk' , \vq} \sum_{l,m_{1} , m_{2} , n} \gre(\vk_1 , i\ep_{m_{1}-})\gre(\vk_2 , i\ep_{m_{2}})\gre(\vk' , i\ep_{n-})\gre(\vk' , i\ep_{n}) \\
            &\times
            \gre(\vk_{1} - \vq , i\ep_{m_{1}} - i\omega_{l})\gre(\vk_{2} - \vq , i\ep_{m_{2}} - i\omega_{l})
            \gre(\vk' - \vq , i\ep_{n} - i\omega_{l}) U(\vq , i\omega_{l-})U(\vq,i\omega_{l}) \\
            &\times  J^{1}_{x}(\vk') J^{2,e-e}_{x}\pqty{\vk_{1} - \frac{\vq}{2}}.
        \end{split}
    \end{equation}
Using the same approximations used in the derivation of Eq. \eqref{eq:phi_drag}, we obtain
    \begin{equation}
        \label{eq:phi_drag_m_e-e_k}
        \begin{split}
            \Phi^{12}_{e-e} (\hbar\omega + i\delta) &= \frac{4\pi^2 i \hbar \omega }{\Gamma V^4}\sum_{\vk_1, \vk_2, \vk' , \vq} N'(\Delta \ep')\pqty{f(\ep_{\vk_1}) - f(\ep_{\vk_1 - \vq})}\pqty{f(\ep_{\vk_2}) - f(\ep_{\vk_2 - \vq})}\pqty{f(\ep_{\vk'}) - f(\ep_{\vk' - \vq})}\\
            &\times \delta(\Delta\ep_{1} - \Delta\ep')\delta(\Delta\ep_{2} - \Delta\ep') U^{R}(\vq,\Delta\ep')U^{A}(\vq,\Delta\ep')J^{1}_{x}(\vk') J^{2,e-e}_{x}\pqty{\vk_{1} - \frac{\vq}{2}} + O(\omega^2).
        \end{split}
    \end{equation}
In the limit of $\Delta \ep_{1} = \Delta \ep_{2} = \Delta \ep' \to 0$, this term disappears, whereas Eq. \eqref{eq:phi_drag} approaches a constant. Therefore, we can conclude that the contributions from Fig.~\ref{fig:drag_ee} is smaller than that in Fig.~\ref{fig:drag}\subref{subfig:drag}.

\end{widetext}

\bibliographystyle{apsrev4-2}
\bibliography{reference}

\end{document}